\documentstyle[12pt]{article}
\newcommand{\functs}{{\cal O}_2}
\newcommand{\supcon}{{\cal SC}}
\newcommand{\k}{{\cal K}}
\title{Old and New Fields on Super Riemann Surfaces}
\author{Jeffrey M. Rabin \\ 
Dept. of Mathematics, U.C.S.D.\\
La Jolla, CA 92093\\
jrabin@ucsd.edu}
\begin{document}
\maketitle
\begin{abstract}
The ``new fields" or ``superconformal functions" 
on $N=1$ super Riemann surfaces introduced recently by Rogers 
and Langer are shown to coincide with the Abelian differentials 
(plus constants), 
viewed as a subset of the functions on the associated 
$N=2$ super Riemann surface.
We confirm that, as originally defined, they do not form a super vector space.
\end{abstract}

It has been known for some time that the globally defined holomorphic functions
on a generic super Riemann surface (SRS) with odd spin structure do not form a 
super vector space \cite{nelson}.
Neither do the holomorphic superconformal tensors of weights $-1$ through $2$
\cite{luke}.
That is, one cannot uniquely express an arbitrary tensor as a linear
combination of some small basis set.
The proposed ``basis" inevitably contains nilpotent tensors which are 
annihilated by some elements of the set of constant Grassmann parameters 
(exterior algebra) $\Lambda$.
Therefore linear combinations with coefficients from $\Lambda$ cannot be unique.
This is a special case of the general fact that sheaf cohomology groups 
$H^i(M,\cal F)$ of complex supermanifolds over $\Lambda$ are $\Lambda$-modules,
but these modules are not necessarily free.
In comparison with the theory of Riemann surfaces used in 
bosonic string theory, this complicates the 
discussion of period matrices, Jacobian varieties, 
functional determinants, and other
ingredients in the construction of superstring amplitudes \cite{dhoker}.
 
Recently, Rogers and Langer introduced a notion of ``superconformal functions"
on a SRS with the intent of overcoming these difficulties \cite{rogers}.
These are a certain subset $\supcon$ of the holomorphic sections of a 
rank $(1,1)$ super vector bundle $E$ over the SRS.
On the basis of a direct but very complicated calculation, the module of global
superconformal functions $H^0(M,\supcon)$ was claimed to be free,
that is, a super vector space.
Subsequently, an error in the calculation was corrected, with the result that
$H^0(M,\supcon)$ is not free but a free submodule can be picked out by
imposing an additional condition on the sections \cite{corrig}.
In this paper we will give a geometric interpretation of $\supcon$ 
leading to a simpler verification of these facts.

One can associate to every $N=1$ SRS, indeed to every  
complex supermanifold $M$ of dimension $(1,1)$, or ``supercurve", 
an $N=2$ (untwisted) SRS $M_2$ \cite{schwarz}.
This correspondence is generically two-to-one, so that each supercurve 
$M$ has a dual supercurve $\hat{M}$ with the same associated $M_2$.
SRSs can be characterized by the property of self-duality: $M=\hat{M}$ iff
$M$ is a SRS \cite{berg}.
We will see that contour integration theory on $N=2$ SRSs \cite{cohn} makes
it possible to integrate Abelian differentials (tensors of weight $1$) 
on $M$ to obtain functions on $M_2$.
The super vector bundle $E$ will be shown to be precisely the sheaf of functions
$\functs$ on $M_2$, and $\supcon$ will be 
that subsheaf obtained by integrating Abelian differentials on $M$.
In fact, $\functs$ splits as a sheaf into the direct sum of the Abelian
differentials $\k$ and the functions $\cal O$ on $M$.
In terms of this splitting, $\supcon$ consists of $\k$ plus integration
constants from $\cal O$.
Since the global Abelian differentials on $M$ do not form a super vector space
\cite{luke,dhoker}, we confirm that the global superconformal functions 
(as originally defined) do not either.

We will be rather casual about supermanifold theory in this paper.
The reader can think of Rogers \cite{oldrogers} or DeWitt \cite{dewitt} 
supermanifolds (topologically vector bundles over their bodies) 
over a finite- or infinite-dimensional complex
Grassmann algebra $\Lambda$ with generators $\eta_i$, 
$i = 1, 2, \ldots, L$ in the finite-dimensional case, or
of families of sheaf-theoretic supermanifolds \cite{sheafrefs} 
over the parameter superspace $({\rm point},\Lambda)$.
All that matters is that the various sheaves and cohomology groups considered are
$\Lambda$-modules; linear combinations are taken with $\Lambda$ coefficients.

\bigskip

Let $M$ be a supercurve, covered by charts $U_\alpha$, with transition functions
\begin{equation}
z_\beta = F_{\beta\alpha}(z_\alpha,\theta_\alpha),\;\;\;\;
\theta_\beta = \Psi_{\beta\alpha}(z_\alpha,\theta_\alpha) \label{Mtrans}
\end{equation}
in $U_\alpha \cap U_\beta$.
The associated split supercurve $M_{\rm split}$ is obtained by setting to zero
the generators $\eta_i$ of $\Lambda$ in the transition functions.
The body or reduced space $M_{\rm red}$ 
is obtained by dropping $\theta$ as well.
We denote by $\k$ the canonical bundle of $M$, the line bundle having
transition functions $k^{-1}_{\beta\alpha}$, where
\begin{equation}
k_{\beta\alpha} = 
{\rm sdet}\left[\begin{array}{cc}
\partial_z F_{\beta\alpha} & \partial_z \Psi_{\beta\alpha} \\
\partial_\theta F_{\beta\alpha} & \partial_\theta \Psi_{\beta\alpha} 
\end{array}\right].
\end{equation}
Its sections will be called Abelian differentials.
The associated $N=2$ SRS $M_2$ has coordinates $(z,\theta,\rho)$ with
the transition functions \cite{schwarz,berg}
\begin{equation}
z_\beta = F_{\beta\alpha}(z_\alpha,\theta_\alpha),\;\;\;\;
\theta_\beta = \Psi_{\beta\alpha}(z_\alpha,\theta_\alpha),\;\;\;\;
\rho_\beta = \rho_\alpha k_{\beta\alpha}(z_\alpha,\theta_\alpha) + 
\phi_{\beta\alpha}(z_\alpha), \label{N=2transfuncts}
\end{equation}
where
\begin{equation}
\phi_{\beta\alpha} = \frac{\partial_\theta F_{\beta\alpha}}
{\partial_\theta \Psi_{\beta\alpha}}.
\end{equation}
The sheaf of functions $\functs$ on $M_2$ is then an extension of $\k$ by the
sheaf $\cal O$ of functions on $M$; there is an exact sequence
\begin{equation}
0 \longrightarrow {\cal O} \stackrel{\rm inc}{\longrightarrow} \functs 
\stackrel{D^+}{\longrightarrow} \k \longrightarrow 0, \label{exact}
\end{equation}
where $D^+ = \partial_\rho$.
That is, the (local) functions on $M_2$ include the functions on $M$
as those which do not depend on $\rho$, whereas the coefficient of $\rho$ in a
function on $M_2$ transforms as a section of $\k$.
The compatibility of the transition functions imposes the following cocycle
relation on $\phi_{\beta\alpha}$:
\begin{equation}
\phi_{\gamma\alpha}(z_\alpha) = 
\phi_{\gamma\beta}(z_\beta) + 
k_{\gamma\beta}(z_\beta,\theta_\beta) \phi_{\beta\alpha}(z_\alpha).
\end{equation}
This is the cocycle in $H^1(M,\k^{-1})$ which classifies extensions of the form
(\ref{exact}).
If a function $G(z,\theta,\rho)$ on $M_2$ is written in the form
$g(z,\theta) - \rho \gamma(z,\theta)$, then the pair $(g,\gamma)$ transforms as
\begin{equation}
 \left[\begin{array}{c}g_\beta \\ \gamma_\beta\end{array}\right] = 
\left[\begin{array}{cc}1 & k^{-1}_{\beta\alpha}\phi_{\beta\alpha} \\
0 & k^{-1}_{\beta\alpha}\end{array}\right]
\left[\begin{array}{c}g_\alpha \\ \gamma_\alpha\end{array}\right],
\label{vectorbundle}
\end{equation}
i.e., as a section of a rank $(1,1)$ vector bundle $E$ over $M$.
(A.S. Schwarz has informed me that the equivalence between $\functs$ and $E$
has been observed previously \cite{icmschwarz}.)
If the extension cocycle $\phi_{\beta\alpha}$ vanishes, 
this reduces to the direct sum of line bundles ${\cal O} \oplus \k$.
If the cocycle is cohomologically trivial, meaning
\begin{equation} \label{cohomtriv}
\phi_{\beta\alpha} = \sigma_\beta - k_{\beta\alpha}\sigma_\alpha
\end{equation}
for some functions $\sigma_\alpha(z_\alpha,\theta_\alpha)$ on $U_\alpha$,
then the change of coordinates 
\begin{equation} \label{coordchange}
\tilde{\rho}_\alpha = \rho_\alpha - \sigma_\alpha(z_\alpha,\theta_\alpha)
\end{equation}
induces a change of basis $(g,\gamma) \rightarrow (g-\sigma\gamma,\gamma)$
in the fibers of $E$ which again reduces it to $\cal O \oplus \k$.
We will show below that this cocycle is always trivial when $M$ is a SRS.

The correspondence between supercurves $M$ and $N=2$ SRSs is generically
two-to-one; there is a dual curve $\hat{M}$ yielding the same $M_2$.
To construct it, introduce in each chart the new coordinate
$u = z - \theta\rho$.
One finds that $(u,\rho)$ then transform independently of $\theta$, so they
patch together as coordinates for another supercurve, $\hat{M}$.
Its transition functions are
\begin{equation}
u_\beta = \hat{F}_{\beta\alpha}(u_\alpha,\rho_\alpha),\;\;\;\;\;\;\;\;
\rho_\beta = \hat{\Psi}_{\beta\alpha}(u_\alpha,\rho_\alpha),
\end{equation}
where
\begin{equation}
\hat{F} = F + \frac{DF}{D\Psi} \Psi,\;\;\;\;\;\;\;
\hat{\Psi} = \frac{DF}{D\Psi},
\end{equation}
and as usual $D = \partial_\theta + \theta \partial_z$.
The exact sequence analogous to (\ref{exact}) is
\begin{equation}
0 \longrightarrow \hat{\cal O} \stackrel{\rm inc}{\longrightarrow} \functs 
\stackrel{D^-}{\longrightarrow} \hat{\k} \longrightarrow 0, \label{hatexact}
\end{equation}
where $D^- = \partial_\theta + \rho \partial_z$.
These operators $D^\pm$ are the usual superconformal derivatives on an $N=2$
SRS, satisfying $(D^\pm)^2 = 0,\;\{D^+,D^-\} = \partial_z$.
However, $(z,\theta,\rho)$ are coordinates adapted to the projection 
$M_2 \rightarrow M$ rather than the customary superconformal coordinates
[which are $(z-\frac{1}{2}\theta\rho,\theta,\rho)$].

Contour integration theory on an $N=2$ SRS takes the following form
\cite{cohn}.
The derivative of a function $G$ is defined to be the pair $D^+G \oplus D^-G$,
which is a section of $\k \oplus \hat{\k}$.
Conversely, an antiderivative $G$ of such a section 
$\omega \oplus \hat{\omega}$ always exists locally and is unique 
up to a constant.
In coordinates, if $G = a + \theta \alpha - \rho (\beta + \theta b)$, then
\begin{eqnarray}
D^+G \oplus D^-G & = & (-\beta - \theta b) \oplus 
[\alpha + \rho (a'+b + \theta \alpha')].      \nonumber     \\        
& = & [-\beta(z) - \theta b(z)] \oplus
[\alpha(u) + \rho a'(u) + \rho b(u)].           \label{derivs}
\end{eqnarray}
In finding an antiderivative, $\alpha,\beta,$ and $b$ can be read off directly,
and the only actual integration required is in finding $a$ from $a'$.
The definite (contour) integral $\int_{P_0}^P \omega \oplus \hat{\omega}$, 
with $P_0$ a fixed
$\Lambda$-valued basepoint and $P$ a variable point of $M_2$, is defined
directly as $G(P) - G(P_0)$ when the contour lies in a single contractible chart,
and in general by subdividing any contour into such pieces.
It is generally a multivalued function of $P$, 
changing by periods when $P$ is taken around homology cycles.
This integration theory on $M_2$ is consistent with the corresponding 
integration theories on $M$ and $\hat{M}$ \cite{berg}.
If $c_2$ is a contour on $M_2$, it projects down to contours $c$ and $\hat{c}$
on $M$ and $\hat{M}$ respectively, and we have the relation,
\begin{equation}
 \int_{c_2} \omega \oplus \hat{\omega} =  \int_{\hat{c}} \omega + 
\int_c \hat{\omega}.
\end{equation}
(Note that differentials on a supercurve are invariantly integrated over
contours on the dual curve.)
In particular, for closed contours, $c_2,c,\hat{c}$ all define the same
homology class and the periods of $\omega,\hat{\omega}$ sum to the periods
of the pair.

Now let us specialize to the case that $M$ is itself an $N=1$ SRS
(with odd spin structure) \cite{crane}.
Then $M$ is self-dual; $M=\hat{M}$ \cite{schwarz} (and in fact {\it only}
SRSs are self-dual \cite{berg}).
Its transition functions are superconformal, and conventionally parametrized as
\begin{equation}
z_\beta = F_{\beta\alpha}(z_\alpha,\theta_\alpha) = 
f_{\beta\alpha}(z_\alpha) + \theta_\alpha \psi_{\beta\alpha}(z_\alpha) 
\sqrt{f'_{\beta\alpha}(z_\alpha)},
\label{srstrans}
\end{equation}
\begin{equation} \label{oddtrans}
\theta_\beta = \Psi_{\beta\alpha}(z_\alpha,\theta_\alpha) = 
\psi_{\beta\alpha}(z_\alpha) + \theta_\alpha 
\sqrt{f'_{\beta\alpha}(z_\alpha) + 
\psi_{\beta\alpha}(z_\alpha)\psi'_{\beta\alpha}(z_\alpha)}.
\end{equation}
The canonical bundle $\k$ has the transition function
\begin{equation}
k^{-1}_{\beta\alpha} = (D_\alpha \Psi_{\beta\alpha})^{-1},
\end{equation}
satisfying $D_\beta = k^{-1}_{\beta\alpha} D_\alpha.$
This is the usual bundle of superconformal tensors of weight $1$.
$M_2$ has the additional transition function
\begin{eqnarray}
\rho_\beta & = & \psi_{\beta\alpha}(z_\alpha) + \rho_\alpha 
[\sqrt{f'_{\beta\alpha}(z_\alpha) + 
\psi_{\beta\alpha}(z_\alpha)\psi'_{\beta\alpha}(z_\alpha)} + 
\theta_\alpha \psi'_{\beta\alpha}]   \nonumber  \\
& = & \psi_{\beta\alpha}(z_\alpha) + \rho_\alpha k_{\beta\alpha}.
\end{eqnarray}
The cocycle $\phi_{\beta\alpha}$ characterizing $\functs$ as an extension of $\k$
is simply $\psi_{\beta\alpha}$.
Then the vector bundle description (\ref{vectorbundle}) of $\functs$ precisely
reduces to the rank $(1,1)$ vector bundle $E$ on $M$ defined by Rogers and Langer, 
cf. Eq. (11) of \cite{rogers}.

Because $M$ is self-dual, any Abelian differential $\omega$ on $M$ is also
a differential on $\hat{M}$, so it gives rise to a pair $\omega \oplus -\omega$
which can be integrated on $M_2$.
Furthermore, all periods of this pair vanish:
\begin{equation}
 \oint_{c_2} \omega \oplus -\omega = \oint_c \omega - \oint_c \omega = 0.
\end{equation}
Thus, once we choose a basepoint $P_0$, each section $\omega$ of $\k$ maps to
a single-valued global function on $M_2$ by
\begin{equation}
 \omega \mapsto \int_{P_0}^P \omega \oplus -\omega,
\end{equation}
and a change of basepoint alters this function by a constant.
In coordinates, 
\begin{eqnarray}
\omega \oplus -\omega & = & [-\beta(z) - \theta b(z)] \oplus 
[\beta(u) + \rho b(u)]  \nonumber    \\
& = & (-\beta - \theta b) \oplus [\beta + \rho (b + \theta \beta')].
\end{eqnarray}
Compared to the general case (\ref{derivs}) 
we have $\alpha = \beta$ and $a'=0$.
The antiderivative is trivially
\begin{equation}
G(z,\theta,\rho) = a + \theta \beta(z) - \rho[\beta(z) + \theta b(z)],
\end{equation}
with $a$ constant.

A superconformal function is defined by Rogers and Langer as a (global) 
section of $E$ locally represented by pairs $(g,\gamma)$ satisfying 
\begin{equation}
\gamma(z,\theta) = D[g(z,\theta) + r(z)],
\end{equation}
for some $r(z)$.
This is equivalent to the constraint $\alpha = \beta$ on the
corresponding section $G$ of $\functs$.
The global superconformal functions automatically 
satisfy the constraint $a'=0$ as well \cite{rogers}, so we have identified them 
as the subset of the functions on
$M_2$ obtained by integrating global sections of $\k$.
They are in one-to-one correspondence with such sections, 
apart from an integration constant.

The local situation becomes clear after a change of coordinates 
(\ref{coordchange}) that reduces $\functs$ to a sum of line bundles.
Indeed, condition (\ref{cohomtriv}) for the cohomological triviality of
$\psi_{\beta\alpha}$ is satisfied with $\sigma_\alpha = \theta_\alpha$
as an immediate consequence of (\ref{oddtrans}).
The transition function for the new coordinate $\tilde{\rho} = \rho - \theta$
is then
\begin{equation}
\tilde{\rho}_\beta = \tilde{\rho}_\alpha k_{\beta\alpha}.
\end{equation}
Writing functions on $M_2$ in the form 
$a + \theta \tilde{\alpha} - \tilde{\rho} (\beta + \theta b)$,
where $\tilde{\alpha} = \alpha - \beta$,
explicitly shows the splitting $\functs = {\cal O} \oplus \k$.
Although Rogers and Langer consider superconformal functions only as global
objects, we can also view them locally as sections of the subsheaf $\supcon$ of
$\functs$ defined by $\tilde{\alpha}=0$ and $a=$ constant in every chart.
Indeed, the second condition follows from the first for infinite-dimensional
$\Lambda$, since the preservation of $\tilde{\alpha}=0$
under arbitrary coordinate changes
(\ref{srstrans}) requires $\psi a' = 0$ for all odd $\psi$, which makes $a$
constant in this case (for finite-dimensional $\Lambda$ we
can add functions proportional to $\eta_1 \eta_2 \cdots \eta_L$).
In terms of the splitting $\functs = {\cal O} \oplus \k$, $\supcon$ consists
of $\k$ plus constants from $\cal O$.

The global sections of $\k$ have been discussed in \cite{luke} and 
completely described in \cite{dhoker}.
For the split SRS $M_{\rm split}$, 
sections have the form $\omega = \beta + \theta b$,
with $\beta$ a holomorphic spinor on the body $M_{\rm red}$
and $b$ an Abelian differential on the body.
One can choose a basis of such spinors and differentials and try to modify
the resulting split sections $\omega$ order by order in the generators of 
$\Lambda$ to obtain a basis of sections on $M$.
In \cite{dhoker}
this problem is analyzed in the 2d supergravity picture
with the $\Lambda$-dependence of the transition functions of $M$ 
(i.e., the supermoduli) described by a super Beltrami differential $\chi$,
with the result that $\beta$ can always be extended to a section, 
but $\theta b$ extends iff $b$
obeys a nilpotent constraint (Eq. (6.23) of \cite{dhoker})
$\int d^2z\; h(z) \chi(z) b(z) = 0$ for every
holomorphic spinor $h$ (generically there is exactly one up to
normalization) on $M_{\rm red}$.
The usual Abelian differentials $b$ will generally not satisfy this constraint,
but since $\chi$ is odd, multiples by sufficiently 
nilpotent elements of $\Lambda$ will, 
showing that the global sections $H^0(M,\k)$ are not freely generated.
As a result, neither are the superconformal functions $H^0(M,\supcon)$.
The extra condition imposed on superconformal functions in \cite{corrig}
has the effect of restricting them to the free submodule generated by the
extensions of the spinors $\beta$ only.

We have seen that, for any SRS $M$, $\functs$ splits as
${\cal O} \oplus \k$ because the extension cocycle
$\psi_{\beta\alpha}$ is trivial in $H^1(M,\k^{-1})$.
The reader may still wonder how this is possible 
without the supermoduli which distinguish $M$ from $M_{\rm split}$ 
(more precisely, since the bosonic moduli in
$f_{\beta\alpha}$ may also contain nilpotents, from a projected SRS
having $\psi_{\beta\alpha}=0$) being trivial as well.
The two issues are conceptually distinct: the splitting of $\functs$ means that
a change of the $\rho$ coordinate can remove the term $\psi$ from its 
transition functions, while the projectedness of $M$ would mean that a 
superconformal change of the $(z,\theta)$ coordinates could remove $\psi$ from
their transition functions.
The obstructions to the latter lie in the tangent space to supermoduli space,
$H^1(M,\k^{-2})$ \cite{lebrun}.
To avoid discussion of the global structure of supermoduli space, consider
the issue at the linearized level only.
To lowest order in the generators $\eta_i$ of $\Lambda$, sections of 
$\k^{-1}$ and $\k^{-2}$ split into components as
\begin{equation}
\k^{-1} = \k^{-1}_{\rm red} \oplus \theta \k^0_{\rm red},\;\;\;\;
 \k^{-2} = \k^{-2}_{\rm red} \oplus \theta \k^{-1}_{\rm red},
\end{equation}
where $\k^{-1}_{\rm red}$ is the square root of the holomorphic tangent
bundle to $M_{\rm red}$.
To this order, $\psi = \eta_i \psi_i$ with $\psi_i$ belonging to
$H^1(M_{\rm red},\k_{\rm red}^{-1})$, so that as we know $\psi$ is a cocycle in
$H^1(M,\k^{-1})$ (in fact, to all orders in the $\eta_i$), while 
$\theta \psi$ is a cocycle in $H^1(M,\k^{-2})$ (to lowest order only).
However, $H^1(M,\k^{-2})$, Serre dual to $H^0(M,\k^3)$, is freely generated
while $H^1(M,\k^{-1})$, dual to $H^0(M,\k^2)$, is not \cite{luke}.
This means precisely that a nilpotent combination of nontrivial classes
$\psi_i$ can be trivial in $H^1(M,\k^{-1})$ but not in $H^1(M,\k^{-2})$.

The question of which classes of supercurves enjoy the splitting
$\functs = {\cal O} \oplus \k$ is discussed further in 
\cite{berg}, particularly for the ``generic SKP curves" which yield 
algebro-geometric solutions to the super KP hierarchies.
In contrast to the situation just discussed for SRSs, for such curves
this splitting is equivalent to projectedness. 

\section*{Acknowledgements}
This work grew out of a collaboration with Maarten Bergvelt \cite{berg}, and
I have enjoyed discussions with him and with Mitchell Rothstein.

\end{document}